\documentclass[journal]{IEEEtran}  
\usepackage[utf8]{inputenc}
\usepackage{amsmath}
\usepackage{blindtext}
\usepackage{hyperref}
\usepackage{multirow}
\usepackage[cmintegrals]{newtxmath}
\usepackage[inline]{enumitem}
\usepackage{graphicx}
\usepackage{siunitx}
\usepackage{caption}
\usepackage{subcaption}
\usepackage{balance}
\usepackage{placeins}

\usepackage{listings}
\usepackage{xcolor}

\definecolor{codegreen}{rgb}{0,0.6,0}
\definecolor{codegray}{rgb}{0.5,0.5,0.5}
\definecolor{codepurple}{rgb}{0.58,0,0.82}
\definecolor{backcolour}{rgb}{0.95,0.95,0.92}

\lstdefinestyle{mystyle}{
    backgroundcolor=\color{backcolour},   
    commentstyle=\color{codegreen},
    keywordstyle=\color{magenta},
    numberstyle=\tiny\color{codegray},
    stringstyle=\color{codepurple},
    basicstyle=\ttfamily\footnotesize,
    breakatwhitespace=false,         
    breaklines=true,                 
    captionpos=b,                    
    keepspaces=true,                 
    numbers=left,                    
    numbersep=5pt,                  
    showspaces=false,                
    showstringspaces=false,
    showtabs=false,                  
    tabsize=2
}

\title{WiFiEye -- Seeing over WiFi Made 
Accessible}
\author{\IEEEauthorblockN{Philipp H. Kindt \IEEEauthorrefmark{1},
Cristian Turetta\IEEEauthorrefmark{2}, Florenc Demrozi\IEEEauthorrefmark{2},\\
Alejandro Masrur\IEEEauthorrefmark{1},
Graziano Pravadelli\IEEEauthorrefmark{2},
Samarjit Chakraborty\IEEEauthorrefmark{3}
\\
\IEEEauthorrefmark{1}TU Chemnitz, Germany,\\
\IEEEauthorrefmark{2}University of Verona, Italy,\\
\IEEEauthorrefmark{3}University of North Carolina at Chapel Hill, USA}}
\date{March 2022}

\begin{document}
	This is the author's draft of an article to be subitted to an IEEE publication.
	 Copyright may be transferred without notice, after which this version may no longer be accessible.
	 A copyright notice will be added here upon submission, and additional information in the case of an acceptance/publication.
\bstctlcite{IEEEexample:BSTcontrol}
    \maketitle
\newif \ifcomment

\begin{abstract}
While commonly used for communication purposes, an increasing number of recent studies consider WiFi for sensing. 
In particular, wireless signals are altered (e.g., reflected and attenuated) by the human body and objects in the environment. This can be perceived by an observer to infer information on human activities or changes in the environment and, hence, to ``see'' over WiFi. 
Until now, works on WiFi-based sensing have resulted in a set of custom software tools -- each designed for a specific purpose. Moreover, given how scattered the literature is, it is difficult to even identify all steps/functions necessary to build a basic system for WiFi-based sensing. This has led to a high entry barrier, hindering further research in this area. There has been no effort to integrate these tools or to build a {\em general software framework} that can serve as the basis for further research, e.g., on using machine learning to interpret the altered WiFi signals. To address this issue, in this paper, we propose {\em WiFiEye} -- a generic software framework that makes all necessary steps/functions available ``out of the box''. This way, WiFiEye allows researchers to easily bootstrap new WiFi-based sensing applications, thereby, focusing on research rather than on implementation aspects. To illustrate WiFiEye's workflow, we present a case study on WiFi-based human activity recognition.   

\end{abstract}

\section{Introduction}
\noindent\textbf{``Seeing'' over WiFi:} Most of us rely on WiFi for communication purposes, e.g., for connecting our smartphones and laptops with each other and to the Internet. 
This has catapulted WiFi to ubiquity over the last years and, hence, WiFi signals are almost everywhere. While propagating from a sender to a receiver, WiFi signals are altered by the human body and/or objects in the environment. 
An external observer can perceive these changes in the received signals and, e.g., detect the presence of humans and/or their activities.
Hence, beyond communication, WiFi networks can actually be used to sense or ``see''~\cite{ma:19}. 
Recently, the \textit{IEEE 802.11bf Task Group} was created with the goal of standardizing WiFi-based sensing in the future, which underpins the growing importance of this area~\cite{restuccia2021ieee}.

Applications of WiFi-based sensing are as varied as interesting. 
For example, the number of subjects in a room can be estimated~\cite{demrozi2021estimating}, or their activities can be classified, detecting whether somebody walks, lies/rests, or falls onto the floor~\cite{wang:15}. 

Compared to other technologies, WiFi-based sensing has a number of decisive advantages. 
First, no sensors need to be worn on the body. 
Second, WiFi networks are anyway deployed for communication purposes. As a result, WiFi-based sensing is more economical than most other wireless approaches, e.g., Ultra-Wideband (UWB) or radar-based ones.
Third, WiFi-based sensing is intrinsically privacy-preserving, since it reveals significantly less personal data than, for example, camera-based approaches.

\begin{figure}[!t]
    \centering
    \includegraphics[width=0.9\columnwidth,page={1}]{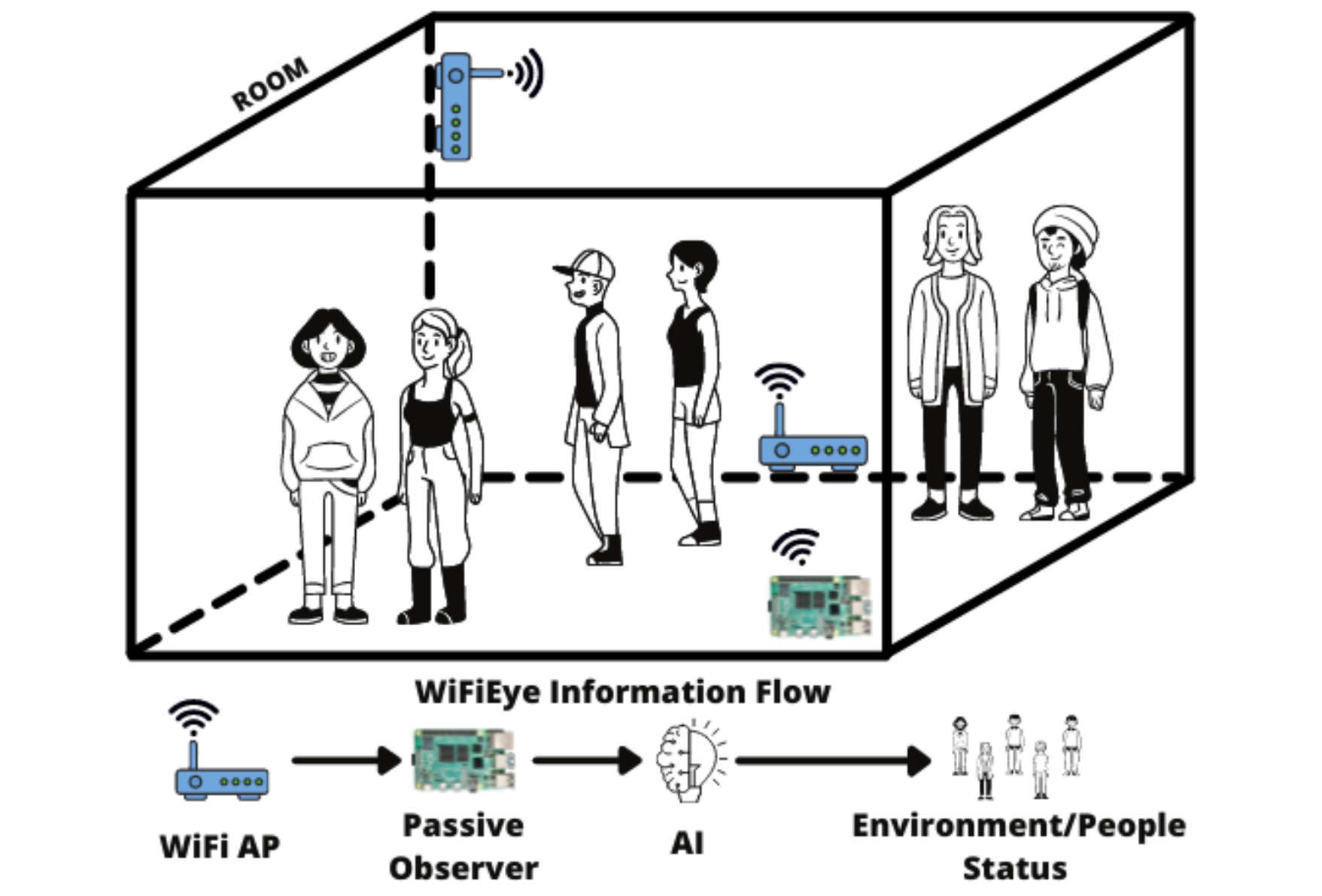}
    \caption{A typical scenario for WiFi-based sensing, consisting of at least one AP and an observer that collects CSI data from the environment. This data is then analyzed with the help of ML methods.}\label{fig:fig_1}
\end{figure}

Figure~\ref{fig:fig_1} shows a typical scenario involving WiFi-based sensing. It consists of one or more access points (APs), a passive observer, and one or more human subjects performing different activities. 
The passive observer extracts the \textit{channel state information} (CSI) from the captured packets, which we describe in more detail in Section~\ref{sec:background}. 
The CSI data is then analyzed to infer information on the activities of these subjects. 
Typically, this is done using machine learning (ML) methods, e.g., support vector machines (SVMs), decision trees (DTs), or convolutional neural networks (CNNs). 
For example, a human subject carries out multiple different gestures in a WiFi environment. For each gesture, CSI data is recorded and fed into a ML model for training.
Once trained, such a model can reliably classify unknown CSI data, thereby achieving an accuracy of more than $\SI{90}{\percent}$~\cite{ma:19}.\\

\noindent\textbf{Setting up a base system:} 
Advances in CSI data analysis are hindered by the high effort involved in creating a working base system. In particular, a workflow for WiFi-based sensing consists of data acquisition and recording, format conversion, data preprocessing, visualization, and classification. All of these steps need to be realized before any research on the actual classification, e.g., using machine learning methods, can be carried out. 

In more detail, first, it is necessary to ``unlock'' CSI data by flashing a custom firmware onto the WiFi radio. 
This is mandatory because no off-the-shelf radio firmware grants access to the CSI data to an external program running on the host computer.

Second, CSI data cannot be fed directly into a ML framework, such as Tensorflow or PyTorch. Usually, multiple parameterizable, application-specific preprocessing, filtering and data format conversion steps need to be applied before the data becomes usable. 

Third, determining the positions of the senders (i.e., APs) and the observer that best suit the intended application requires experience and intuition. Other aspects to be optimized are the set of APs to be considered, the parameter values of different preprocessing steps, and the traffic characteristics of the network. Typically, this optimization is carried out iteratively by repeatedly modifying the setup, recording a sequence of CSI data and analyzing the impact of the modifications on the captured CSI data, e.g., using multiple plots. Multiple iterations are necessary until a suitable configuration is found, which is a time-consuming and cumbersome procedure. The currently used iterative offline-workflow is hence not very effective. Instead, if the preprocessed CSI data could be visualized in \textit{real-time}, the impact of every change could be observed immediately, leading to better results with a reduced effort. 

Unfortunately, existing software tools for WiFi-based sensing only provide a subset of the functions needed in this workflow.
As a result, before developing and optimizing any ML techniques, \emph{significant effort has to be invested into obtaining a working base system}. In addition, to the best of our knowledge, though being of tremendous use, none of the existing software tools support real-time visualization and an online-adjustment of all parameter values.\\

\noindent\textbf{Contributions:} 
This work aims to substantially reduce the initial effort for developing a WiFi-based sensing system. To this end, we introduce a software tool called \textit{WiFiEye} and make it available to the public under an open-source license\footnote{Available at \url{https://github.com/pkindt/WiFiEye}}. 
It provides a ready-to-use base workflow, such that downstream WiFi-based sensing applications can focus on the machine learning aspects. WiFiEye, for the first time, provides a graphical, intuitive, and interactive real-time workflow, which greatly facilitates the development and analysis of WiFi-based sensing systems.
By presenting WiFiEye, we make the following contributions:
\begin{enumerate}
    \item We compile the knowledge from the related literature into ready-to-use software.
    \item We provide an adjustable, working base-system ``out of the box'', thereby minimizing the effort when developing new WiFi-based sensing applications.
    \item We propose a system to process and visualize all data in real-time for supporting the optimization of the setup and its parameter values interactively, and to support running a classifier in real-time.
\end{enumerate}

WiFiEye works together with a Raspberry Pi using a Nexmon-based firmware~\cite{gringoli:19, nexmon:project}, as well as literally any ML framework, e.g., Tensorflow, Keras or PyTorch. Using WiFiEye, all required steps including data recording, format conversion, preprocessing, and visualization, are already implemented and ready-to-use. The entire workflow can be parameterized and optimized at runtime, while visualizing the effects of every adjustment in real-time. Most processing steps are carried out via plugins, which can be activated, configured and deactivated independently from each other. Additional plugins can be developed by everyone with negligible effort. Thereby, the functionality of WiFiEye can be extended or modified easily for custom projects.
Researchers can use WiFiEye as a working and adjustable base system and focus on the actual data classification tasks. By publishing WiFiEye, we intend to foster further innovations and new applications in this area, making this topic more accessible for both the networking as well as the ML community.\\

\section{Current State of WiFi-Based Sensing Practice}
\label{sec:background}
In this section, we first provide the necessary background on WiFi-based sensing and then describe a commonly used workflow and its shortcomings.

\subsection{Background: Channel State Information (CSI)}
When a wireless signal $X$ is sent, the signal $Y$ received by another device is given by 
\begin{equation}
    \mathbf{Y} = \mathbf{H} \circ \mathbf{X} + \mathbf{N},
\end{equation}
where $N$ is related to noise, whereas $H$ represents the CSI. Here, $\circ$ represents an element-wise multiplication.
The 802.11ac protocol supports channels of $\SI{20}{MHz}$ to $\SI{160}{MHz}$ bandwidth.
Using Orthogonal Frequency Division Multiplexing (OFDM), each channel is subdivided into $N=64$ (for $\SI{20}{MHz}$ channels) to $N=512$ (for $\SI{160}{MHz}$ channels) subcarriers, and data is transmitted simultaneously on all of them, except for a few guard subcarriers. 
Due to the small bandwidth, the channel for each subcarrier is assumed to be \textit{flat}, i.e., all frequencies of the subcarrier undergo approximately the same perturbations. Hence, $\mathbf{H} = [h_1, h_2,..., h_N]$ for a received WiFi Frame is a vector of complex values $h_i$, where the index $i$ identifies the subcarrier. 
The magnitude of every complex number $h_i$ represents the attenuation the signal has undergone, whereas the phase expresses the phase change induced, e.g., by human motion~\cite{demrozi2021estimating}. 

Using Automatic Gain Control (AGC), every WiFi receiver continuously adjusts the CSI magnitude on average over all subcarriers to lie within a certain range. Hence, the measured CSI relates to momentary changes in attenuation, while persistent changes are cancelled out by AGC. In contrast, another signal called the \textit{Received Signal Strength Indicator (RSSI)} accounts for the average received power over all subcarriers after antenna and cable loss, but before AGC takes effect. The RSSI is typically an integer that indicates the received power in $dBm$.
We next describe the steps involved in a commonly used workflow for WiFi-based sensing and its shortcomings.

\subsection{Workflow for WiFi-Based Sensing}
As of today, most workflows for WiFi-based sensing research function as follows.\\

\noindent\textbf{Gathering CSI:} 
The firmware of off-the-shelf WiFi radios does not expose the CSI data to any external software. But for a few selected WiFi SoCs, e.g., some instances of the recent Broadcom \textit{BCM43xx} device family, as well as the older Intel \textit{IWL5300} radio and some Atheros SoCs, custom firmware is available. For example, research projects such as Nexmon~\cite{halperin:11, nexmon:project} provide dedicated firmware patches for the recent BCM4355c0 SoC used in a Raspberry Pi, which enables reading the CSI and RSSI.

The patched radio firmware provides the CSI data in the form of a local UDP broadcast on the Raspberry Pi. These UDP packets are usually written into a file on the Raspberry Pi using standard command line tools, e.g., \textit{tcpdump}. 
Such files are then manually copied to a PC or laptop for further analysis.
The format of the CSI data extracted from the \textit{PCAP} files depends on the actual chip of the BMC43xx family. A combination of dedicated libraries (e.g., \textit{libpcap} and CSIKit~\cite{forbes:20}) are required to extract the data. Preprocessing and (offline-)visualization is either carried out by custom developments in environments like MATLAB, or are partially provided by dedicated libraries such as CSIKit~\cite{forbes:20}.\\

\noindent\textbf{Preprocessing:} Next, multiple preprocessing steps on the data are required. First, the raw data from the UDP packets needs to be split up into amplitude and phase. Moreover, parts of the data correspond to \textit{guard subcarriers} and hence do not carry any information on the environment. They need to be removed from the data stream. Additionally, the phase data contains jumps of $2\times \pi$, which need to be removed. Furthermore, the attenuation of the wireless signal caused by the environment is available in the form of the \textit{RSSI} and the \textit{CSI}, as described earlier. It is useful to merge CSI and RSSI into a frequency-dependent aggregated attenuation~\cite{gao:21}. In addition, CSI and RSSI data is often filtered, e.g., by Lowpass, Hampel or Wavelet filters~\cite{schaefer:21}. The actual set of required preprocessing steps depends on the individual application. While some steps can be reused from standard software or dedicated CSI tools, most of them are left over to be implemented by every developer on their own. \\

\noindent\textbf{Setup Optimization:} After an initial system has been developed, the setup needs to be optimized. For example, how sensitive the CSI data is to a certain event of interest depends on the positions of the APs and the receiver, which need to be optimized empirically. Other parameters that can potentially be tuned are the set of considered APs, the wireless channels to be used, and the traffic characteristics of the network.
Furthermore, the filters used for data preprocessing have multiple tunable parameters that impact the results, e.g., coefficients for low-pass filtering the RSSI signal. These parameters also need to be optimized empirically for a particular scenario. For this purpose,  CSI data is first recorded and then visualized for analysis. Based on the results, the setup is then altered and another optimization iteration begins by recording another sequence of CSI data.

CSI data cannot be visualized well using standard tools. First, it has 3 dimensions (i.e. time, subcarrier and actual value). Furthermore, every CSI value is a complex number, and therefore, expands to an amplitude and phase. The amplitude typically exhibits a large dynamic range, and the visualization needs to focus on those ranges of values that represent the events of interest. Since which range of values contains information related to the events of interest is not known a-priori, the visualization itself needs to be adjustable. 
Hence, also for visualization, it is necessary to develop custom solutions or at least to combine and adjust different available tools. 
In addition, repeatedly recording CSI data into files, copying them on a PC or laptop, preprocessing and visualizing them and adjusting the setup after each iteration is time-consuming and cumbersome.\\

\noindent \textbf{Model Generation:} After the setup has been optimized and parameterized, a dataset is usually recorded for the purpose of training a ML model. Furthermore, the data needs to be converted to a format supported by the ML framework, e.g., a comma-separated value (\textit{.csv}) format.

The purpose of the model resulting from the training procedure is to classify previously unseen data in real-time. 
Since a real-time workflow is unavailable in most setups, the recorded dataset is commonly split into testing and training data. The model is trained using the training data and evaluated based on the testing data to mimic an actual online classification. 

\begin{figure*}[!tbh]
    \centering
    \includegraphics[width=1.6\columnwidth,page={2}]{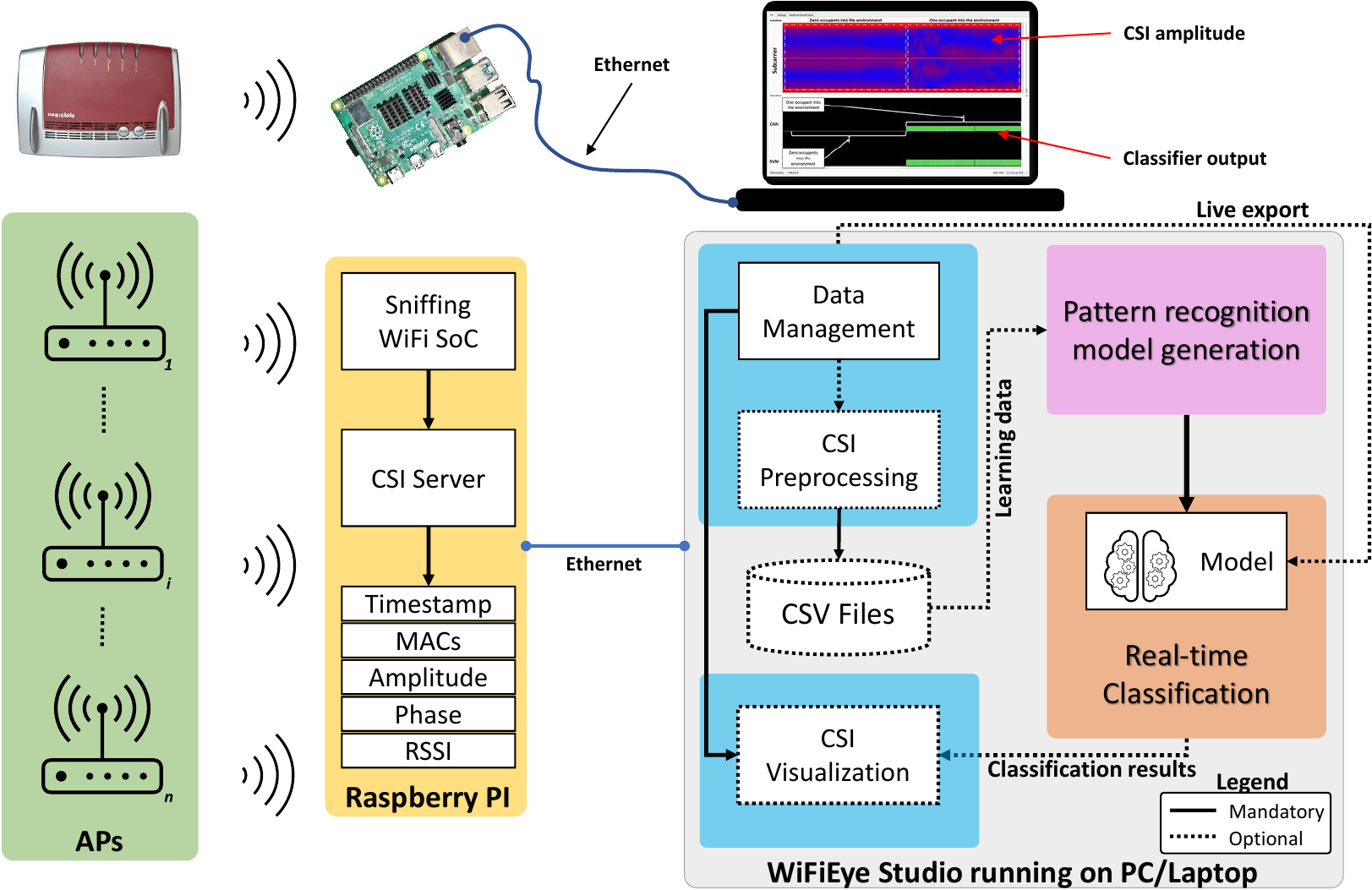}
    \caption{WiFiEye's workflow for WiFi-based sensing.}
    \label{fig:overview}
\end{figure*}

\subsection{Shortcomings}
The workflow described above has multiple significant shortcomings.
First, the entire chain from data acquisition, interpretation and format conversion, preprocessing, model generation to data classification involves a large number of different steps. Some of them are provided by different available tools, whereas others need to be written by every developer on their own. This requires understanding and combining bits and pieces from the literature, from standard software and from dedicated WiFi-based sensing tools, while missing parts require time-consuming custom implementations. Clearly, despite the existence of a dedicated radio firmware that makes CSI data available, significant effort and specialized knowledge is required for developing a working base system.

In addition, all currently available CSI processing tools, to the best of our knowledge, do not support any real-time operation and visualization. As a workaround, the data is recorded on the Raspberry Pi,  manually copied to a PC or laptop and analyzed offline. After adjusting the setup, e.g., by repositioning the passive receiver, another iteration of recording and analysis begins. Repeatedly carrying out this analysis and optimization based on this offline workflow is impractical, since a large number of iterations are needed. This is cumbersome and time-consuming, while the limited number of iterations often does not fully utilize the potential for optimizations of the setup. In contrast, carrying out all processing steps in real-time, adjusting all parameter values online, and studying the impact of any change of the setup in real-time could significantly facilitate the optimization procedure. It could also lead to better results. 

Furthermore, since no tool to deliver preprocessed data to a classification model in real-time is available, the vast majority of works, e.g. \cite{demrozi2021estimating, won:18, Zhang:17,turetta:21}, elude the effort in creating such an online workflow. They instead subdivide previously recorded CSI data into testing and training data. Typically, the testing data is composed of multiple data windows distributed over the entire recorded data. 
The resulting model is evaluated based on this testing data to mimic an actual online classification. The accuracy of the classifier obtained in this way can differ from what would be achieved in an actual online classification, since the environment might change after training the model. A realistic evaluation would again require a workflow that conveys data to a classifier in real-time.

Creating a real-time workflow would require streaming the data from the Reaspberry Pi onto a PC or laptop, performing all preprocessing in real-time, visualizing the CSI data also in real-time, and finally streaming this data into a classifier. Implementing such a real-time workflow further increases the effort that needs to be spent until any research on the machine-learning model can be carried out. 

WiFiEye addresses these shortcomings and drastically reduces the development effort by providing a real-time workflow out-of-the-box. In particular, it provides a complete, pre-configured, parameterizable default workflow, which can be modified and extended easily through a plugin architecture. All processing is done in real-time, and a powerful visualization is provided to also show the preprocessed data in real-time. Hence, the setup can be optimized using \emph{immediate} feedback rather than relying on multiple offline iterations. This helps in selecting the set of APs considered, optimizing the positions of the receiver and the APs, determining the parameter values of the preoprocessing steps, and other optimizations. Finally, WiFiEye supports streaming data to a classifier in real-time, while importing back its classification results, which can be annotated in the real-time visualization. 

\section{The WiFiEye Framework}\label{sec:sensing}
In this section, we first provide an overview on WiFiEye and then describe its components in more detail.

\subsection{Overview}
Figure~\ref{fig:overview} shows a WiFi-based sensing setup involving WiFiEye. 
The green block on the left represents multiple WiFi access points (APs) that emit WiFi signals. For example, every AP continually broadcasts \textit{beacon frames} to advertise its presence with a frequency of around $\SI{9}{Hz}$.
The yellow block in the middle represents a CSI observer, e.g., a Raspberry Pi, which captures WiFi signals and extracts CSI and RSSI data. 
The grey block on the right represents a laptop or PC, which performs signal processing, classification, and visualization. The data recorded into files is used to train a pattern recognition model. This model can be queried in real-time using WiFiEye's live export feature.

\begin{figure*}[bth]
\centering
\includegraphics[width=\textwidth,page={6}]{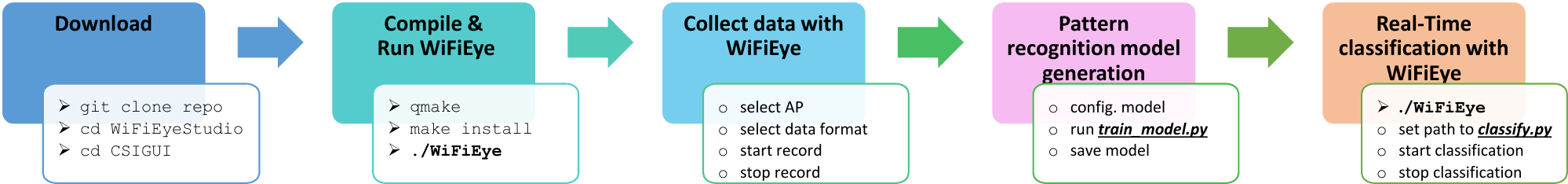}
\caption{Steps required to create a new CSI-based recognition system using WiFiEye.}
\label{fig:steps}
\end{figure*}

Figure \ref{fig:steps} summarizes the steps that need to be carried out to implement an arbitrary, CSI-based pattern recognition scenario using WiFiEye based on a previously prepared Raspberry Pi.
After downloading (1st block), compiling, and running WiFiEye (2nd block), one can collect data by using WiFiEye (3rd block).

To facilitate the setup of an initial ML model, WiFiEye provides two preconfigured python scripts making use of the popular TensorFlow~\cite{tensorflow:15} ML framework. 
A generic, CNN-based pattern recognition model is generated through the \textit{train\_model.py} script (4th block), which contains multiple configuration options. 
It generates a standard CNN architecture, which can be trained for different purposes. 
The model is composed of two \textit{convolutional layers}, which carry out feature extraction based on a time window of CSI data. 
Next, a \textit{max pooling layer} reduces the dimension of the feature map.
After that, two \textit{dense fully connected layers} map the flattened features to the output classes and also compute a confidence value for each of them.
The \textit{classify.py} script (5th block) can be executed from within \textit{WiFiEye Studio} to query the model in real-time.
These two scripts serve as a starting point for developing own models, which are not limited to the described architecture or the Tensorflow library.

\subsection{Data Flow}
The radio firmware extracts CSI data from the \textit{frame preambles} it captures, which carry synchronization information that remains unencrypted even in encrypted networks.
WiFiEye provides a program called the \textit{CSI Server} for the Raspberry Pi. It reads the CSI data from the patched WiFi SoC, timestamps it and transfers it to a PC or laptop in real-time, where it is processed further.

\subsection{WiFiEye's Components}
WiFiEye consists of two software packages. The \textit{CSI Server} is deployed on the Raspberry Pi, whereas an intuitive graphical tool called \textit{WiFiEye Studio} is usually run on a PC or laptop (cf. Figure~\ref{fig:overview}). 
After receiving the data form the \textit{CSI Server}, \textit{WiFiEye Studio} mainly carries out the following 3 tasks:
\begin{enumerate}
    \item Data management: The CSI data is needed at different sinks, and each of them requires different data formats and preprocessing steps. WiFiEye provides all functionalities to control and convert this dataflow.
     \item Preprocessing: Different preprocessing steps are needed to bring the data into a suitable form for analysis, e.g., gain compensation and guard carrier removal/nulling. WiFiEye provides an extensible and adjustable preprocessing architecture. A common workflow is preconfigured to work out-of-the-box, which is described later. 
    \item Visualization: WiFiEye provides a real-time visualization of the CSI amplitude, phase and RSSI data.
    WiFiEye can additionally annotate the output of a classifier in real-time.
    This is helpful for optimizing the setup, tuning the parameters of the preprocessing steps, and analyzing whether a particular setup is sensitive to a certain event of interest.
\end{enumerate}
We next describe these functionalities in more detail. 

\subsubsection{Data Management}
The CSI data can be recorded into data files, exported to a classifier and visualized in real-time. 
WiFiEye provides different preprocessing steps and data formats for each of these possibilities.
For example, it might be beneficial to only visualize data belonging to certain transmitters of interest, while at the same time exporting the data belonging to \emph{all} transmitters into a \textit{.csv} file.
WiFiEye internally converts the data into the required format and allows filtering for one or multiple transmitter MAC addresses, which can be chosen during runtime.
WiFiEye supports recording files in three different formats, i.e., one comma-separated value (\textit{.csv}) format optimized for simplicity, one \textit{.csv} format optimized for size, and a binary format for minimalistic memory requirements.

In addition to exporting data to a file, WiFiEye can launch an arbitrary executable from the graphical user interface. The CSI data is then streamed to the executable by writing to its standard input. Classification results are imported into WiFiEye by reading from the standard output of the executable. 
\subsubsection{CSI Preprocessing}
\label{sec:preprocessing}
The data received from the patched WiFi SoC needs to be preprocessed before recording and real-time export. WiFiEye provides a plugin-based, extensible architecture for this purpose. In particular, each preprocessing step 
is realized by a plugin. Plugins are linked dynamically during runtime, can be compiled individually and are easy to implement. 

\begin{figure}[bth]
\centering
\includegraphics[width=0.95\columnwidth,page={3}]{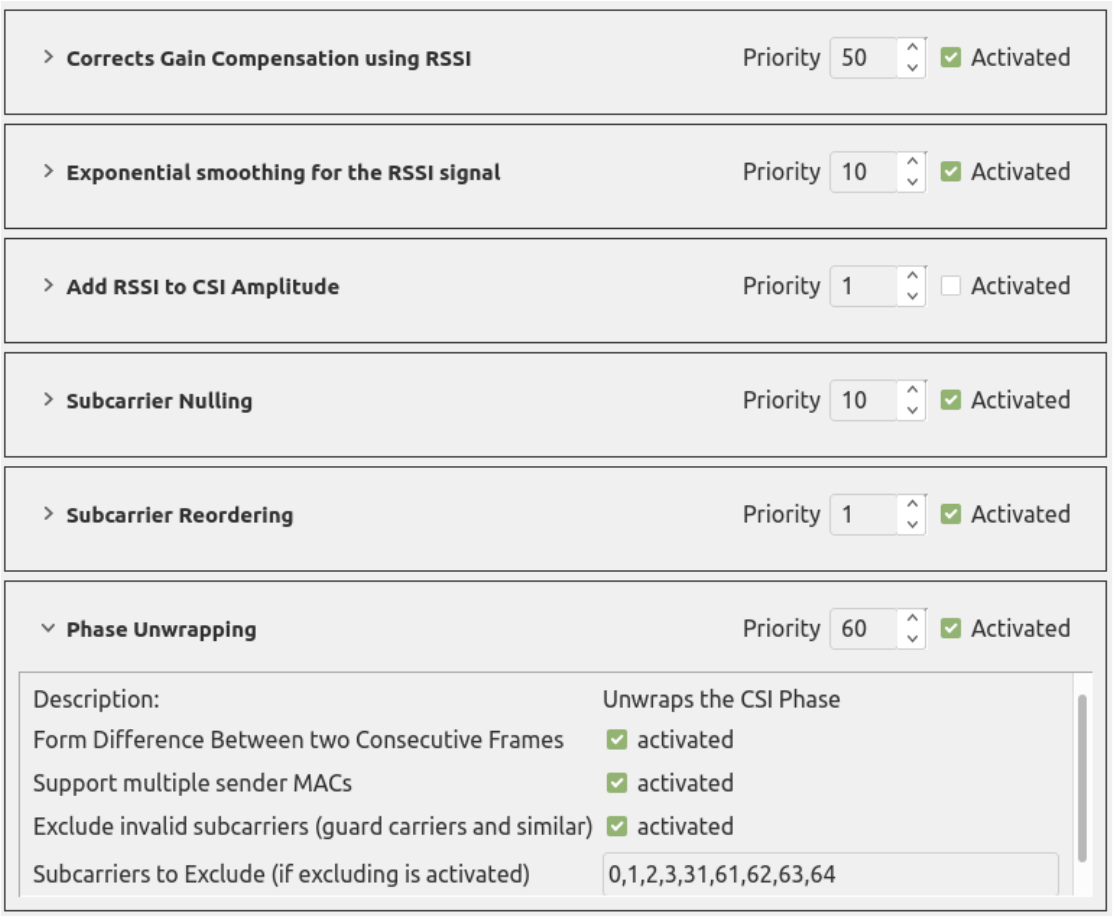}
\caption{Configuring filter plugins through the graphical user interface.}
\label{fig:filters_screenshot}
\end{figure}

As shown in Figure~\ref{fig:filters_screenshot}, WiFiEye Studio allows the user to select the set of active plugins and to determine their execution order by assigning individual priorities. 

Furthermore, every plugin can expose multiple adjustable parameters to the user.

 WiFiEye comes with a comprehensive set of processing plugins. It thereby supports a generic and useful preprocessing workflow, which is described next.\\

 \noindent\textbf{MAC Filtering:} When capturing  WiFi frames, data from all APs in range is received, of which only a few are relevant.
 WiFiEye allows selecting suitable APs interactively during runtime.
\begin{figure*}[htb]
    \centering
     \includegraphics[width=\textwidth]{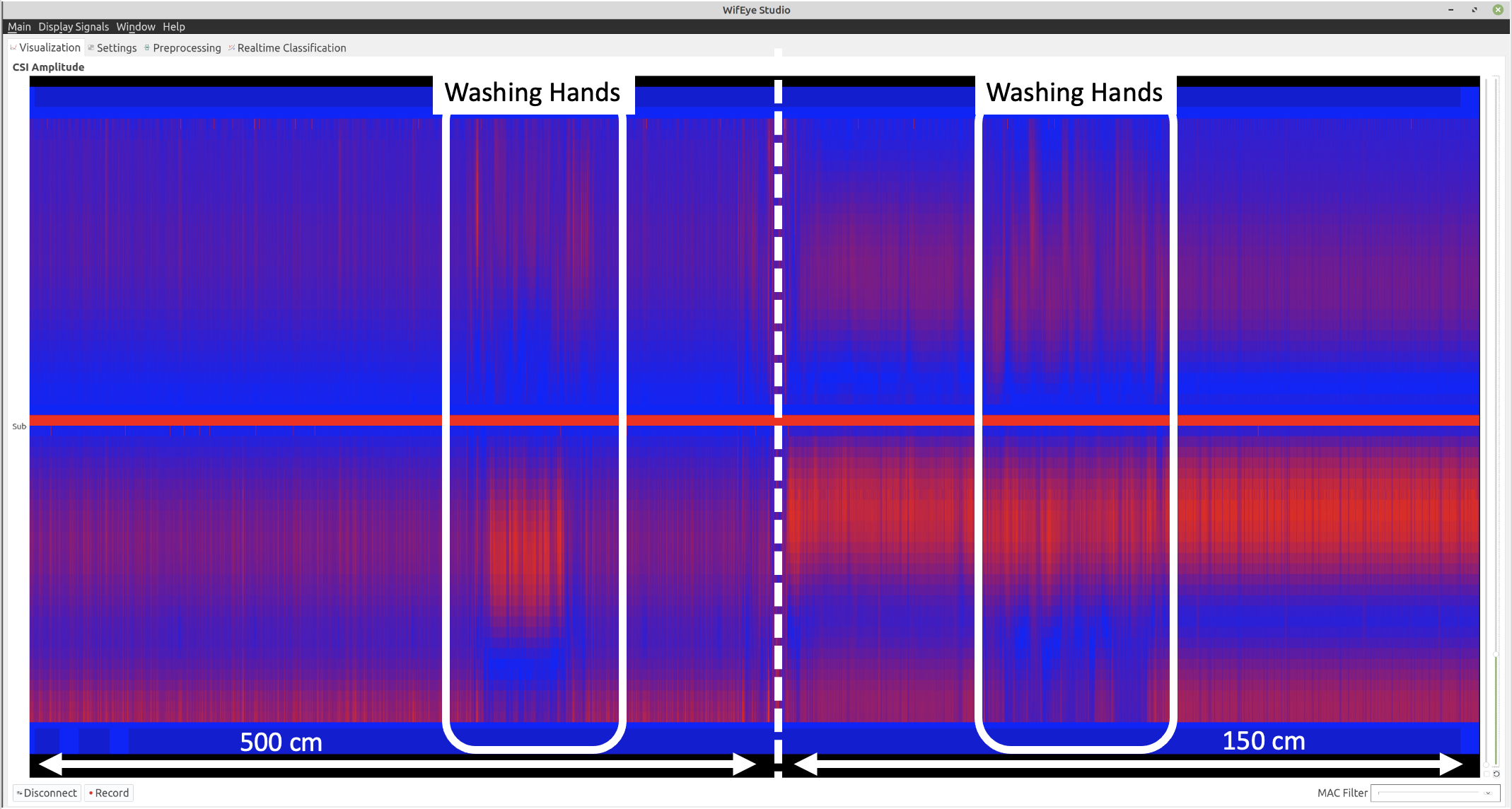}
    \caption{Real-Time visualization of the CSI amplitude. As can be seen, a setup in which the receiver had a distance of 1.5 m from a pair of hands being washed showed a more distinct and fine-grained pattern in the CSI data when somebody made a hand-washing gesture than at a distance of 5 m.}
    \label{fig:visualization_screenshot}
\end{figure*}

\noindent\textbf{Amplitude/Phase
Extraction:} From the complex values $h_i$, $i =1...n$, WiFiEye extracts the amplitudes $a_i$ and phases $\Phi_i$, which can then be processed separately.

\noindent\textbf{Bandwidth Narrowing:} WiFi uses channels with different bandwidths, and the BCM43455c0 WiFi radio on the Raspberry Pi can monitor channels from 20~MHz to 80~MHz bandwidth. When a transmitter only uses e.g., 20 MHz, whereas the radio captures using 80~MHz, most of the data does not contain any usable information. 
WiFiEye supports narrowing the bandwidth of its input data, thereby supporting a mode of operation in which the configuration on the Raspberry Pi never needs to be changed. 
  
 \noindent\textbf{Subcarrier Removal:}
 WiFiEye can set the values of subcarriers that do not carry any valid information to zero to avoid disturbing the ML model.\\
\noindent\textbf{Subcarrier Re-Ordering:} The CSI values on the different subcarriers received from the WiFi SoC are sometimes not ordered in a linear manner.WiFiEye reorders them if needed.\\
\noindent\textbf{AGC Compensation:} WiFi SoCs use automatic gain control (AGC) to bring the signal amplitude and hence the CSI into a desired range. WiFiEye can cancel the impact of AGC using the method described in~\cite{gao:21}.

\noindent\textbf{Denoising the RSSI:} In our experiments, we found that the RSSI is prone to noise to a much larger extent than the CSI. WiFiEye hence provides an exponential smoothing method to denoise the RSSI, thereby also improving the AGC compensation.

\noindent\textbf{Phase Unwrapping and Cleansing:} The resulting phase information from the WiFi SoC contains multiple discontinuities caused by the phase ``wrapping around'' for values greater than $2 \cdot \pi$ or lower than $0$. WiFiEye linearizes the phase data accordingly. 

\subsubsection{CSI Visualization}
Visualizing CSI data in real-time is crucial for optimizing the setup.
WiFiEye can simultaneously visualize the CSI amplitude, CSI phase and RSSI in real-time, as shown in Figure~\ref{fig:visualization_screenshot}.

CSI amplitude and phase are displayed using two independent, synchronized, 3-dimensional plots. One dimension is given by the reception time of the frame. The second dimension is the subcarrier index, and the third dimension is formed by the CSI amplitude or phase.
In WiFiEye, time forms the X-axis, the subcarrier index the Y-axis and the CSI amplitude or phase is represented by different colors. Low amplitudes are drawn in blue, high ones in red, while blending from blue to red for intermediate values. 
The depicted time window is updated in real-time by scrolling from the right to the left. 

On the right side of the plot (cf. Figure~\ref{fig:visualization_screenshot}), there are two sliders, which can be used to select two thresholds that define a certain range of CSI amplitudes or phases to be visualized. All values that lie outside of these ranges are discarded, whereas the remaining ones are re-scaled to cover the entire range of possible colors.
Hence, the sliders can be used to selectively visualize the ranges of values that are most sensitive to a certain event or activity.

For visualizing the output of a classifier, a bar is painted (see Figure~\ref{fig:overview}). The height of each bar corresponds to the class number to which the considered time window of CSI data has been assigned, and the bar color depicts the classification confidence. 
Jointly examining CSI data and classification results allows studying for which kind of input the classifier creates which output.

\section{Case Study}\label{sec:eval}
This section describes a real-world experiment using WiFiEye to collect data, train a pattern recognition model and perform real-time classification.

\subsection{Experimental Setup}
\begin{figure}[h]
\centering
\includegraphics[width=0.9\columnwidth,page={5}]{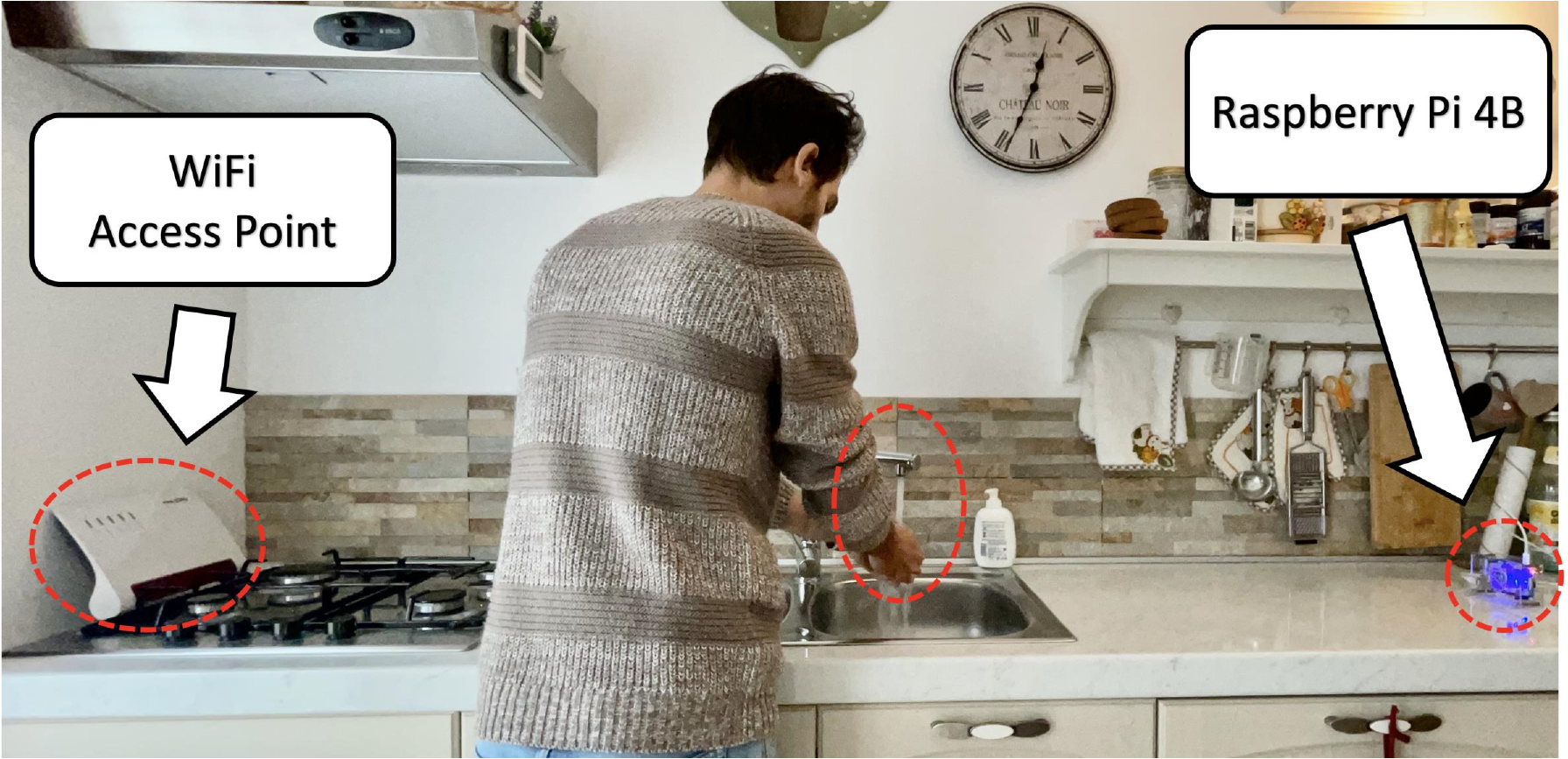}
\caption{Experimental setup.}
\label{fig:scenario}
\end{figure}

In a kitchen, we placed a wireless router as a data source to the left-hand side of a sink, while a \emph{Raspberry Pi 4B} as a WiFi observer was placed on its right-hand side, as shown in Figure~\ref{fig:scenario}. 
The goal was to recognize different human activities, i.e., washing, drying and soaping hands, as well as the sink being unused.

We recorded 5 minutes of CSI data for each of the four different activities and generated a ML model with negligible effort using the \textit{train\_model.py} script provided along with WiFiEye.
To study the effect of systematically optimizing the setup, we have further repeated the experiment for 2 different distances of the Raspberry Pi from the sink, i.e., 150 cm and 500 cm. Figure~\ref{fig:visualization_screenshot} depicts the CSI amplitude for both distances when washing hands. The pattern at a distance of 1.5~m is more distinct and fine-grained as for a distance of 5~m. Our results will show that this is also reflected in the performance of the classifier.

\subsection{Evaluation}
We used $75 \%$ of the recorded data for training our model and $25 \%$ for evaluating the accuracy of the resulting classification, which is based on 1-second time windows. 
For each window, we compared the results with the actually performed activity. We evaluated the performance in terms of the F-score, which is a standard metric that combines precision and recall.

For a distance of the receiver from the sink of 5 m, the classifier could distinguish between these activities with a F-score of 0.93. More precisely, we could correctly classify the sink being idle in 78~\% of all corresponding time windows, soaping hands in 99~\%. washing hands in 100~\%, and drying hands in 92~\% of them. For a distance of 1.5 m, the F-score improved to 0.97. Here, the sink being idle could be correctly classified in 97~\%,  soaping hands with 97~\%, washing hands with 99~\%, and drying hands with 96~\%. 

\section{Concluding Remarks}
\label{sec:concl}
WiFi-based sensing is being actively studied by the community and is expected to further grow in importance in the near future. However, the progress is slowed down by the considerable effort it takes to develop a custom base system. We aim to significantly reduce this effort by introducing WiFiEye, thereby making WiFi-based sensing more accessible by providing the entire workflow 
``out of the box''. In the future, we plan to add support for additional hardware platforms by providing customized versions of WiFiEye's CSI Server for them.

    \balance
    
    \bibliographystyle{IEEEtran}
    \bibliography{IEEEabrv,biblio}{}
\end{document}